\documentclass[preprint,12pt,3p]{elsarticle}

\usepackage{amssymb}
\usepackage{hyperref}
\usepackage{color}
\usepackage{amsmath}
\usepackage{subfigure}
\usepackage{todonotes}
\usepackage{appendix}
\usepackage{mathtools}
\usepackage{comment}
\usepackage[super]{nth}
\usepackage[british]{babel}

\newcommand{\beq}{\begin{equation}}
\newcommand{\eeq}{\end{equation}}
\newcommand{\beqs}{\begin{equation*}}
\newcommand{\eeqs}{\end{equation*}}

\newcommand{\grad}{{\nabla}}
\newcommand{\norm}[1]{\left\lVert#1\right\rVert}

\newcommand{\Tth}{\ensuremath{T}}
\newcommand{\fth}{\ensuremath{f}}
\newcommand{\gth}{\ensuremath{g}}
\newcommand{\hth}{\ensuremath{h}}
\newcommand{\cvx}{\ensuremath{\mathrm{cvx}}}
\newcommand{\No}{\ensuremath{\mathrm{No}}}
\newcommand{\act}{\ensuremath{\mathrm{act}}}
\newcommand{\expt}{\ensuremath{\mathbb{E}}}
\DeclareMathOperator*{\arginf}{arg\,inf}
\DeclareMathOperator*{\argsup}{arg\,sup}

\definecolor{pastelorange}{RGB}{233,131,125}
\definecolor{pastelblue}{RGB}{59,134,179}

\journal{Nucl.~Instrum.~Meth.~A}

\begin{document}
\begin{frontmatter}

\title{Transport away your problems: \\ Calibrating stochastic simulations with optimal transport}

\author[a]{Chris Pollard}
\ead{christopher.pollard@physics.ox.ac.uk}
\author[a]{Philipp Windischhofer}
\ead{philipp.windischhofer@physics.ox.ac.uk}
\address[a]{Department of Physics, University of Oxford, \\Keble Road, Oxford, United Kingdom}

\begin{abstract}

  Stochastic simulators are an indispensable tool in many branches of science.
  Often based on first principles, they deliver a series of samples whose
  distribution implicitly defines a probability measure to describe the
  phenomena of interest.
  However, the fidelity of these simulators is not always sufficient for all
  scientific purposes, necessitating the construction of ad-hoc corrections
  to ``calibrate'' the simulation and ensure that its output is a faithful
  representation of reality.
  In this paper, we leverage methods from transportation theory to construct
  such corrections in a systematic way. We use a neural network to compute minimal
  modifications to the individual samples produced by the simulator such that
  the resulting distribution becomes properly calibrated.
  We illustrate the method and its benefits in the context of experimental
  particle physics, where the need for calibrated stochastic simulators is
  particularly pronounced.
  
\end{abstract}

\end{frontmatter}

\section{Introduction}
\label{sec:intro}

Many of the physical sciences heavily rely on complex simulations to describe processes and
phenomena of interest.
One important category of simulation programs, relevant to this paper, is formed by
\textsl{stochastic simulators} \cite{stochastic_simulation_textbook}.
They model systems or processes that involve elements of randomness, i.e.~whose
behaviour is not uniquely determined by the set of initial conditions.

The challenges associated with the construction of such simulators are nicely illustrated
by an example from experimental particle physics.
In scattering experiments, collisions between energetic particles can access information
about the fundamental physical dynamics at very short length scales.
As probes of quantum mechanical---and hence probabilistic---processes, only the distribution
of a large number of these scattering ``events'' carries meaningful information.

To exploit the experimentally collected data to the fullest extent, it is
important to have access to accurate predictions of the expected event distribution.
Such ``event generators'' must incorporate a large number of physical effects
occurring at very different energy scales, requiring methods ranging from
perturbative calculations to parametrised, phenomenological models.
The resulting simulation models generally have a very large number of free
parameters.
Many of these parameters describe well-understood physical processes (typically
at low energy scales), which are not of interest per se, but must be included in
order to faithfully represent physical reality.
Only a small subset directly concerns the fundamental physics of high-energy
scattering reactions, and are thus of most interest to the experimenter. 

Before being applied to the study of novel physical phenomena, it is vital to
ensure that the simulation can accurately model all known contributing effects. 
For the purposes of this ``calibration'', the output of the simulator is
compared to a suitable calibration dataset, usually taken from experiment. 
The simulation model is then adjusted such that discrepancies are largely
removed (While the word ``calibration'' may carry many meanings, depending on
context, in this paper we take it to mean the correction of a simulation to
reproduce some aspect of observed data).
Often, the causes underlying these differences cannot be interpreted in terms of
the original simulation parameters, or it is simply not feasible to do so. 
Therefore, the applied calibration does not typically take the form of a well-motivated
change of the simulation model or its parameter values, but is rather an ad-hoc
modification of its output.

In this paper, we present a very general and flexible way for the calibration of such a stochastic
simulator, and apply this technique to situations that are typical for experimental particle physics.
In Section \ref{sec:sim_calib_hep}, we give a brief overview of the simulation
models and calibration techniques that are used in this field.
Section \ref{sec:formalisation} then formalises the task of calibrating a given simulator.
Section \ref{sec:OT_for_calibrations} introduces our proposed method by rephrasing this
calibration problem in the language of transportation theory, and Section \ref{sec:implementation}
summarises aspects of its practical implementation.
Finally, Section \ref{sec:results} documents several typical example applications.

\paragraph{Reading guide}
Section \ref{sec:sim_calib_hep}, on applications and calibration techniques in particle physics,
has been written specifically for practitioners of the subject, but can be read independently
of the remainder of the paper.
The main body of the manuscript, Sections \ref{sec:formalisation} to \ref{sec:implementation}, presents
our proposed calibration method in a very general way, and should be accessible as well to readers who are
not familiar with particle physics.
The description of the example applications shown in Section \ref{sec:results},
while again inspired by situations from particle physics, should likewise be accessible.

\section{Calibrated simulators in experimental particle physics}
\label{sec:sim_calib_hep}

Particle collider experiments make use of sophisticated collision and
detector simulations that have been refined over many decades~\cite{mcgens, geant4,
atlsim, cmssim}.
While these simulations have been carefully tuned and match observed data remarkably
well, the large datasets coming from machines such as the LHC, for instance, lead to
extremely demanding environments.
It is common, therefore, to calibrate various aspects of these simulation models as part
of the data analysis to remove residual simulation deficiencies.
Quantities in need of calibration often feed rather directly into the final
analysis results, underlining the importance of well-behaved calibration techniques.

\begin{figure}[tph]
  \centering
  \subfigure[][]{
    \includegraphics[width=6cm]{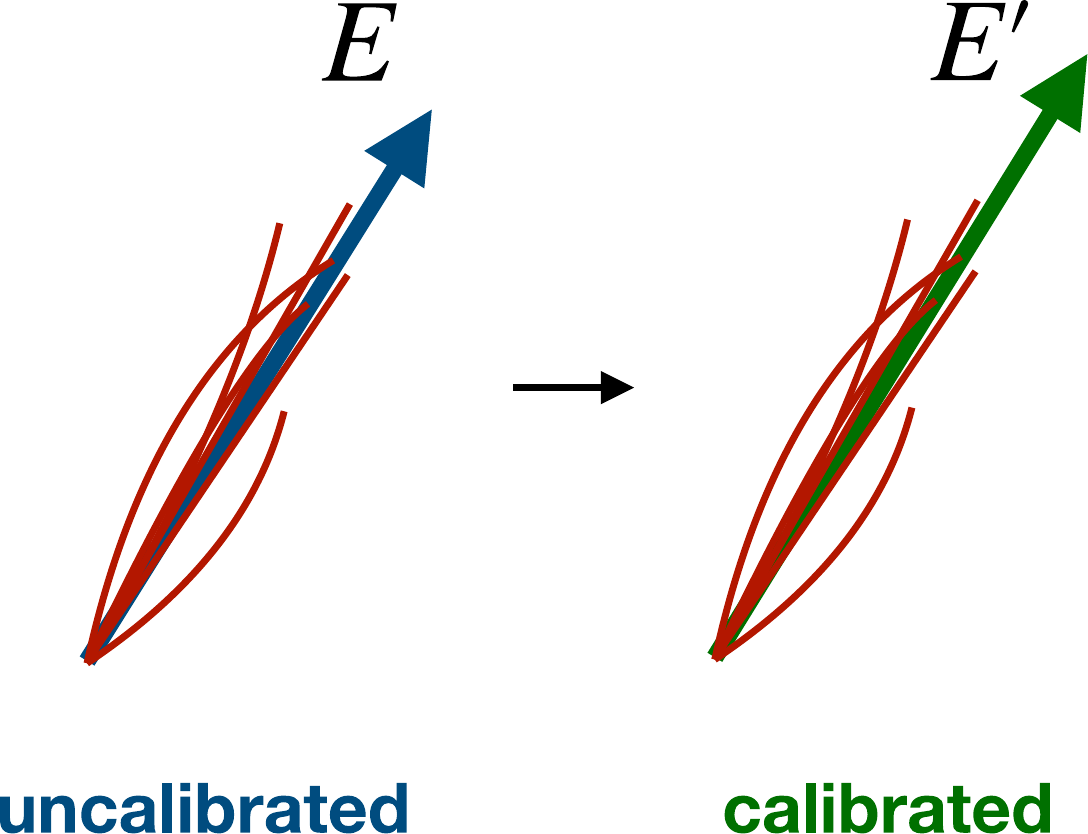}
    \label{subfig:jet_calibration}
  }
  \hspace{2cm}
  \subfigure[][]{
    \includegraphics[width=6cm]{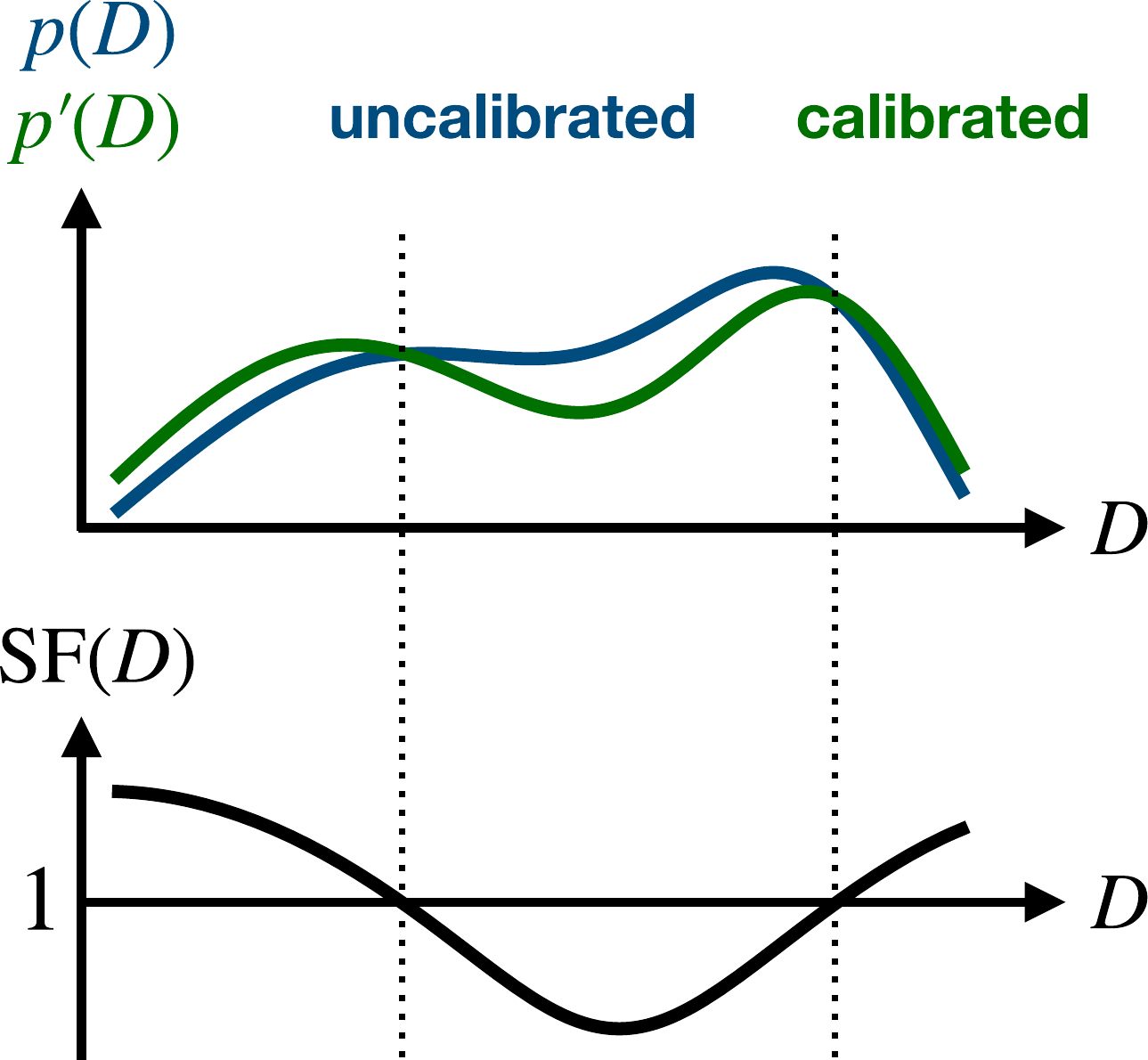}
    \label{subfig:disc_calibration}
  }
  \caption{In \protect\subref{subfig:jet_calibration}, the calibration replaces a jet with
  measured energy $E$ with a jet with a different energy $E'$ in order to
  correct the jet energy scale and resolution.
  In \protect\subref{subfig:disc_calibration}, scale-factors are used to ensure
  proper modelling of the distribution of a discriminant $D$.}
  \label{fig:SF_illustration}
\end{figure}

For example, corrections to the average muon and electron energy responses
in relevant detectors are derived using decays of $Z$ bosons to charged
leptons~\cite{atlegamma, atlmuons, cmsegamma, cmsmuons}, taking advantage of
precision measurements of the $Z$ boson pole mass from previous
experiments~\cite{PDG}.
Similar modifications are derived for the first two moments of the jet energy response
using momentum-balancing techniques at hadron colliders~\cite{atljets, cmsjets}.
The calibration is applied by modifying the respective quantity or object \emph{in-situ},
as shown in Figure \ref{subfig:jet_calibration}.

The LHC experiments also make heavy use of high-level quantities with less-obvious physical
meanings than ``energy'' and ``mass''.
For example, high-performance discriminators, which are built from artificial
neural networks or boosted decision trees and are trained to identify ``interesting''
objects in collision events based on many input signals from the
detector, e.g. for $b$-tagging~\cite{atlbtag, cmsbtag}, are ubiquitous in modern
collider data analyses.
These discriminators involve complex calculations with dozens or hundreds of
input features, and may in turn appear rather opaque.
Correcting the collision simulation for these intricate observables can
be a challenge because the detector response is not very strongly constrained by the
physicist's prior knowledge or intuition---simply correcting the mean or width of
the output distribution is very likely not sufficient, as it is for a distribution that
is approximately gaussian like the lepton or jet energy response.

One common technique for the calibration of such general distributions involves calculating
binned density ratios between collision data and predictions from the simulation for the
distribution of interest, as illustrated in Figure \ref{subfig:disc_calibration}.
This has proven very useful e.g. in the case of $b$-jet tagging~\cite{atlbtag, cmsbtag}.
These density ratios are commonly known as ``scale-factors''; in the calibration procedure,
they act as additional weights associated to the calibrated object.
Scale-factors are derived in well-understood control samples and may be parametrised by
one or more other observables; the transverse momentum $p_{T}$ or the pseudorapidity
$\eta$ are common scale-factor parameters at a hadron collider.

Density ratios built from binned distributions do come with some drawbacks, however.
Bins with no predicted events cannot be corrected, and the scale-factor weights
in general change the total number of predicted events when applied in regions
that do not have the same pre-calibration distribution in simulation. 
In addition, binned density estimates generalise very poorly to multidimensional settings,
which makes this technique unsuitable for calibrations where dependencies on many variables
need to be considered.
Recent developments derive high-dimensional, un-binned scale factors using
boosted decision trees~\cite{Martschei:2012pr}, artificial neural
networks~\cite{Rogozhnikov:2016bdp}, and adversarial neural
networks~\cite{cmsadv}, although the possibility of changing the total number of
predicted events by applying these weights remains.

In the following, we lay out an approach based on transportation theory that
does not suffer from these shortcomings.

\section{Formalisation of the problem}
\label{sec:formalisation}

We now make the notion of ``calibrating'' a stochastic simulator more precise, and set up the
notation that will be used in the following.
For the purposes of this paper, a stochastic simulator is an arbitrary
computational program that implicitly
defines a probability distribution $p(x|\theta)$ by generating samples (or ``events'') $x\in X$
that are distributed according to this distribution, i.e.~$x \sim p(x|\theta)$.
The vector $\theta$ labels the set of free parameters of the simulation, e.g.~physical constants.
(The simulation can also have discrete parameters, which e.g.~select a specific
model from a set of alternatives.)
We shall not impose any restrictions on the form and content of the samples $x$.
In general, $x$ is a high-dimensional vector whose components fully describe the final state
of the simulation.
In certain cases, it can be convenient to allow the simulator to deliver weighted
samples, i.e.~to produce pairs $(x, w)$, where $w$ takes the role of an event weight.

A set of generated samples, $\mathcal{X}=\{(x_i, w_i)\}_{i=1\ldots N}$ allows expectation
values of arbitrary observables to be estimated,
$\langle\mathcal{O}\rangle\approx\sum_{i=1}^N w_i \, \mathcal{O}(x_i) / \sum_{i=1}^N w_i$.
If required, density estimation techniques (such as histograms) give access to an explicit estimate
of the underlying probability distribution $p(x|\theta)$.

To calibrate the simulator, the simulated probability distribution $p(x|\theta)$ is compared
to the distribution $q(y), \,\,y\in Y$ the simulator is intended to sample from.
That is, $q(y)$ describes the \emph{actual} behaviour of a physical system, and not merely
its model.
In practice, $p(x|\theta)$ is usually very close (but not identical) to $q(y)$ for suitable
simulation parameters $\theta$.
As these densities are intractable and not directly accessible, the calibration is performed
by comparing the simulated dataset $\mathcal{X}$ to the calibration dataset
$\mathcal{Y} = \{(y_j, w_j)\}_{j=1\ldots M}$ drawn from $q$.
In many cases $\mathcal{Y}$ is obtained experimentally, which usually results in events $y_j$
that carry unit weight.

To eliminate any residual differences between $p(x|\theta)$ and $q(y)$ (and thus also between
$\mathcal{X}$ and $\mathcal{Y}$), a calibration procedure is applied to the output of the
uncalibrated simulator, i.e.~the dataset $\mathcal{X}$.
Given an event $(x_i, w_i) \in \mathcal{X}$, any admissible calibration procedure thus generates
a modified event $(x_i', w_i')$.
The collection of these events, $\mathcal{X}' = \{(x_i', w_i')\}_{i=1\ldots N}$, is then said
to come from the calibrated simulator, which samples from the calibrated distribution $p'(x'|\theta)$.
For this procedure to be successful, $p'(x'|\theta)$ should be indistinguishable from $q(y)$
for the purposes of computing arbitrary expectation values and estimating densities.
We denote this equivalence as $p'(x'|\theta)\equiv q(y)$.
(In practice, of course, the quality of the calibration may only be evaluated based on the
equivalence of the corresponding collections of samples $\mathcal{X}'$ and $\mathcal{Y}$.)
A precise formulation of this requirement in the context of our proposed calibration method
follows below in Section \ref{sec:OT_for_calibrations}.

\section{Transportation theory for calibrations}
\label{sec:OT_for_calibrations}

The calibration strategies mentioned in Section \ref{sec:sim_calib_hep} belong to two distinct
categories.
Scale factors solely affect modifications to the event weight, i.e.~implement calibration
procedures of the type $(x_i, w_i) \rightarrow (x_i, w_i')$.
Modifications of the to-be-calibrated quantities themselves (such as the jet energy) manifestly
do not change the weight, and thus correspond to a calibration step of the form $(x_i, w_i) \rightarrow (x_i', w_i)$.
In a typical application in particle physics, techniques of both kinds can be present simultaneously
to correct various different aspects of the simulation.

In this section we focus on the latter case, and outline a systematic method to construct
such a calibration for very general situations.
As we shall see, in various special cases, this technique reduces to prescriptions that are already
well-known to practitioners, and thus motivates and formally justifies them.

A general, weight-agnostic calibration method brings with it a number of immediate conceptual benefits.
First, modifying the event directly leads to greater flexibility in the application of the
calibration:
the calibration can easily be restricted to only affect certain components of $x$ (e.g.~those that
are known to be poorly modelled) while ensuring that components for which the simulator is thought
to be accurate are not modified at all.
Second, the normalisation of the dataset is manifestly preserved by the calibration, while this
need not always be the case for weight-based techniques.
Finally, no additional weight variance is induced by the calibration and the statistical power
of the available samples is not diluted.

Here and in the following, we consider only the case where all components of the event
vector $x$ are continuous, and the corresponding probability densities are sufficiently smooth.
This restriction is not strictly needed, and our method can be extended to also apply to
discrete probability distributions, albeit at the expense of added technical complexity.

\subsection{Description of the method}

\begin{figure}
  \centering
  \subfigure[][]{
    \includegraphics[width=6cm]{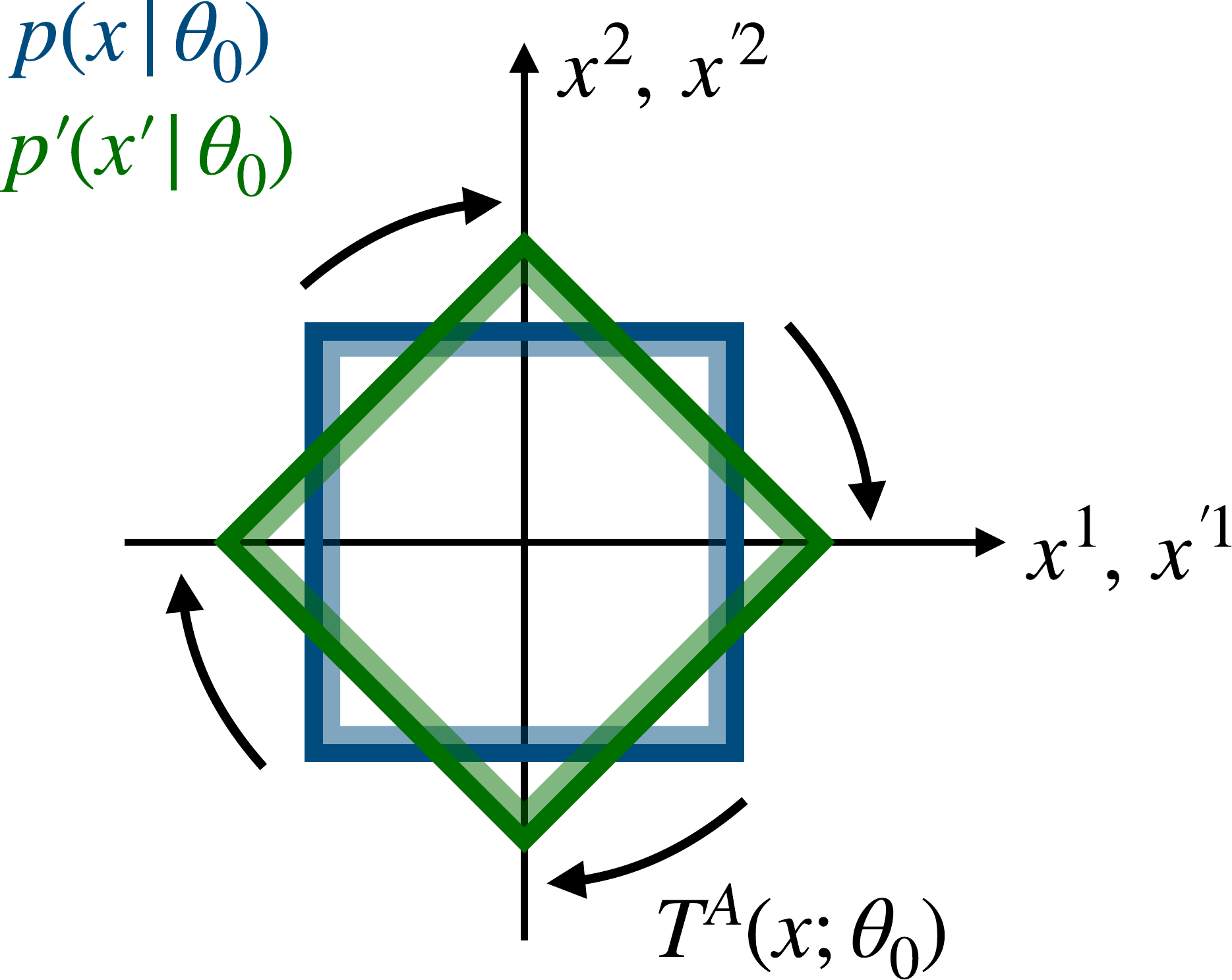}
    \label{subfig:rotation}
  }
  \hspace{2cm}
  \subfigure[][]{
    \includegraphics[width=6cm]{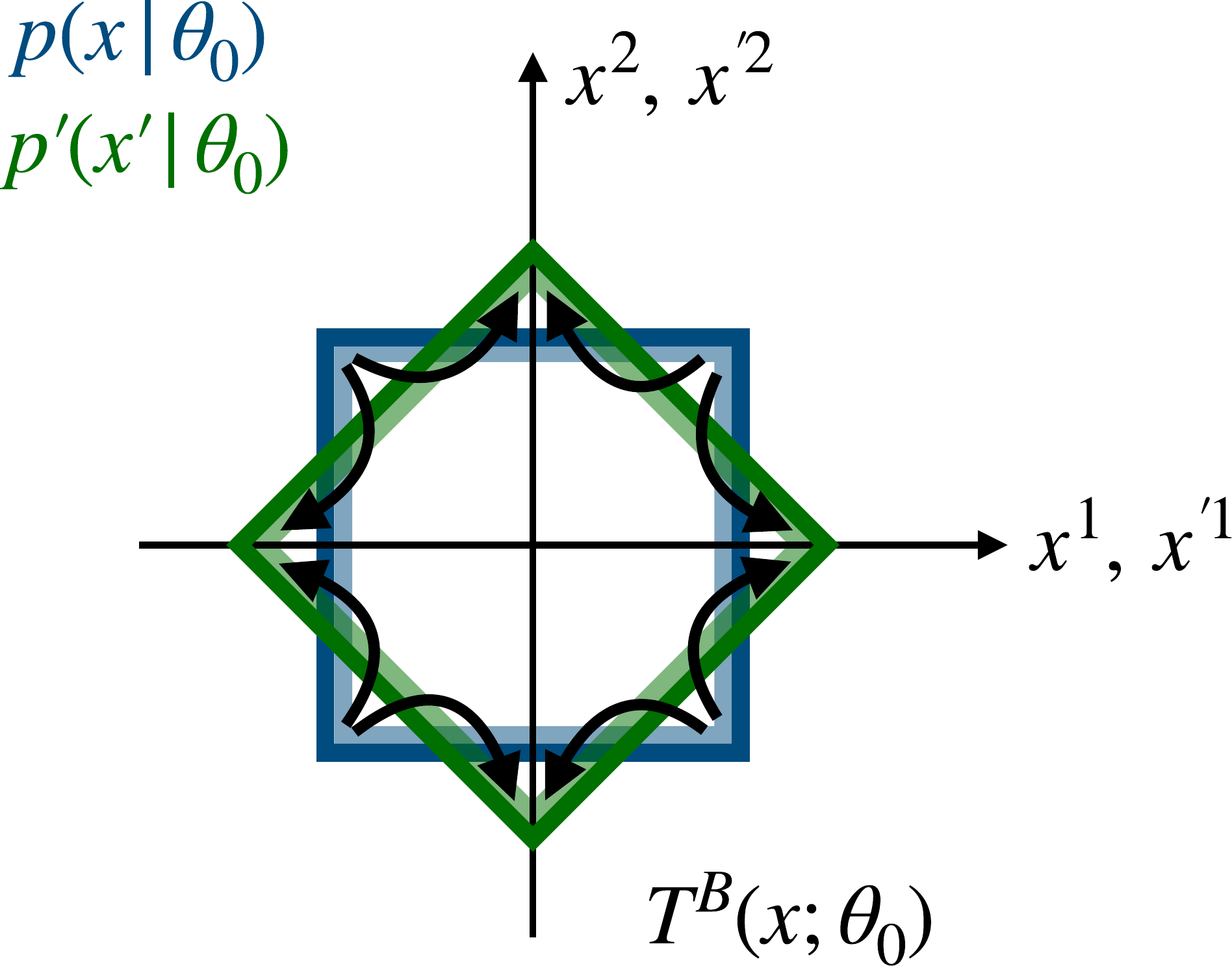}
    \label{subfig:morphing}
  }
  \caption{Events $x=(x^1, x^2)$ produced by the uncalibrated simulator with simulation parameters $\theta_0$ are
    uniformly distributed in the inside of an axis-aligned square (blue), while a properly calibrated
    simulator uniformly populates with events $x'=(x'^1, x'^2)$ a square which is rotated by $\pi/2$ (green).
    The function $T^A(x;\theta_0)$ in \protect\subref{subfig:rotation} achieves a proper calibration by
    \emph{actually} rotating the distribution, while $T^B(x;\theta_0)$ in \protect\subref{subfig:morphing}
    performs a more complicated transformation that leads to the same result.
  }
  \label{fig:cartoon_transport}
\end{figure}

The goal thus becomes to compute a deterministic correction $x' = \Tth(x;\theta)$ for each event
such that the simulator becomes properly calibrated, i.e.~$p'(x'|\theta) \equiv q(y)$.
Note that, since the uncalibrated distribution $p(x|\theta)$ depends on the parameters $\theta$,
but the calibration target $q(y)$ does not, also the function $\Tth$ implementing the
calibration becomes dependent on $\theta$.
The function $\Tth$ need not be unique, i.e.~there can exist nonequivalent
solutions which lead to identical calibrated distributions $p'(x'|\theta)$.
A simple example is shown in Figure \ref{fig:cartoon_transport}.

However, the uncalibrated simulator already contains a significant amount of domain knowledge.
Naturally, the calibration should not negate this structure, but merely fine-tune the simulator
output.
Therefore, one should give preference to the function $\hat{\Tth}(x;\theta)$ that modifies the simulator
to the minimal degree possible, while still achieving a proper calibration.

That is, the preferred solution to the calibration problem is the one with minimal expected
modifications of the simulator,
\beq
\hat{\Tth}(x;\theta) = \arginf_{T(x;\theta): \,\, p'(x'|\theta)\equiv q(y)} \,\, \int dx \, p(x|\theta) \, c\left[x, \Tth(x;\theta)\right],
\label{eq:monge}
\eeq
where the minimum is taken over all functions $\Tth$ which successfully calibrate the simulator.
Written explicitly, the requirement $p'(x'|\theta)\equiv q(y)$ asserts that
\beqs
\int_{\Tth^{-1}(B;\theta)} dx\, p(x|\theta) = \int_B dy\, q(y), \quad \forall B \in Y,
\eeqs
i.e.~that the calibrated distribution $p'(x'|\theta)$ and $q(y)$ allocate equal probability mass
to arbitrary regions $B\in Y$.
The ``cost function'' $c\left[x, T(x;\theta)\right]$ provides a measure for the degree to which an event
is modified by the calibration procedure.
This distance measure should be chosen in accordance with the physics content of the observables that are
being calibrated. It forms an important input to the method that allows the user to encode prior knowledge
and steer the properties of the solutions of Equation \ref{eq:monge}.
For quantities that are inherently unbounded (e.g.~energy differences), the Euclidean
distance might be a reasonable choice. In general, more complicated distance measures might be
necessary to fully capture the content of the participating objects (angular variables are a common
example). This aspect is illustrated below in Section \ref{subsec:towards_solution_quadratic}.
It's interesting to note that similar ideas have been studied recently in high
energy physics to define metrics between collider events~\cite{Komiske:2019fks,
Cai:2020vzx}.

The optimisation problem in Equation \ref{eq:monge} has been intensely studied for more than a
century in the context of transportation theory, where its solution is interpreted as the
optimal way to transport a commodity between locations of production and locations of demand.
In this context, it is also known as the ``Monge optimal transport problem''
\cite{monge_ot}.

The proper analysis of the problem in Equation \ref{eq:monge} requires levels of mathematical rigour
that might obstruct its implementation from a practitioner's perspective, which we prefer to take here.
We thus proceed by briefly sketching some relevant aspects on an intuitive level and refer to the
mathematical literature (see references below) for all details.

\subsection{Towards a solution for the quadratic cost}
\label{subsec:towards_solution_quadratic}

The characteristics of the solution of Equation \ref{eq:monge} turn out to strongly depend on
the form of the cost function $c[x,x']$.
It is easy to see that for the Euclidean cost $c[x,x']=\norm{x-x'}$, many equally ``optimal''
solutions are possible (imagine shifting a row of books to the right by one: one can either move
the entire collection of books by one place to the right, or pick the leftmost book and place
it at the far right \cite{geometry_of_OT}).

A special role is played by the quadratic cost $c[x,x']=\frac{1}{2}\norm{x-x'}^2$.
For this case, the optimal solution to Equation \ref{eq:monge} is unique, and can be
constructed as the gradient of a scalar field, which plays the role of a ``potential''.
Note that this property already excludes the function $T^A(x;\theta_0)$ in Figure \ref{subfig:rotation} as the optimal
solution of the situation shown there: rotations produce nonvanishing curl, while gradient
fields are always curl-free. It turns out that the function $T^B(x;\theta_0)$ in Figure \ref{subfig:morphing}
\emph{does} stem from a potential, and that, moreover, it is indeed the solution that
minimises the expectation of the quadratic cost. The event modifications resulting from $T^B(x;\theta_0)$ contain a
significant radial component. However, if the radial distribution of events is known to be well-modelled by the simulation,
any such radial transport should be avoided. In this example, one can use a distance measure that---unlike the quadratic cost---differentiates
between modifications along the radial and azimuthal directions and that can thus suppress any radial transport%
\footnote{One would generalise the line element of the Euclidean plane, $ds^2 = dr^2 + r^2 d\phi^2$,
  where $r$ is the radial coordinate and $\phi$ the azimuthal angle.
  A weighted line element $ds^2 = \alpha^2 dr^2 + r^2 d\phi^2$ with $\alpha > 1$ would have the desired effect:
  it defines a metric that is conformally equivalent to that of a cone with an opening angle inversely proportional to $\alpha$.
  Indeed, the generalisation of Equation \ref{eq:monge} to Riemannian manifolds has been the subject of intense study \cite{OT_and_curvature},
  and analogues of the results presented here for the case of the Euclidean plane continue to hold.
  The practical solution of the optimal transport problem in this general setting introduces additional technical challenges that are beyond the scope of this paper,
  but that can be addressed e.g.~by using the methods presented in Ref.~\cite{riemannian_convex}.}.
This provides a way to effectively channel corrections to the simulation so that they are compatible with prior expectations.

To construct the optimal solution for the quadratic cost, we make use of the Kantorovich duality relation and Brenier's theorem, both of which are central
and celebrated results in the theory of optimal transportation.
(A detailed and rigorous discussion of these results can be found e.g.~in the monograph in Ref.~\cite{villani_topics}.)
For the present situation, they relate the formulation in Equation \ref{eq:monge} to the
following, ``dual'', formulation of the optimisation problem,
\beq
\hat{\fth}(y;\theta) = \arginf_{\fth \in \cvx(Y)} \int dy \, q(y) \fth(y;\theta) + \int dx \, p(x|\theta) \fth^*(x;\theta).
\label{eq:dual}
\eeq
The minimum in Equation \ref{eq:dual} is taken over the set of functions $\fth(y;\theta)$ that are convex w.r.t.~the argument $y$, which we denote as $f \in \cvx(Y)$.
The function $\fth^*(x;\theta)$ is the convex conjugate of $\fth(y;\theta)$ defined by
\beq
\fth^*(x;\theta) = \sup_{y\in Y} \left[ x \cdot y - \fth(y;\theta) \right].
\label{eq:legendre_transform}
\eeq
It is not immediately obvious that Equations \ref{eq:monge} and \ref{eq:dual} indeed produce identical
solutions.
We refer to Sections 1.1 and 2.1 (in particular Theorem 2.12) of Ref.~\cite{villani_topics} and
also to Ref.~\cite{OT_with_convex_NNs} for a thorough explanation of all steps involved in a
proof of their equivalence.

The optimal transport map $\hat{\Tth}(x;\theta)$ that solves the original problem may be reconstructed from
the function $\hat{\fth}(y;\theta)$ that attains the extremum in the dual problem in Equation \ref{eq:dual},
\beq
\hat{\Tth}(x;\theta) = \grad_x \hat{\fth}^*(x;\theta),
\label{eq:gradient_reconstruction}
\eeq
i.e.~the transport function is a pure gradient field. The function $\grad_y \hat{\fth}(y;\theta)$ also has
an interpretation as a transport function: it solves the ``inverse'' transport problem, i.e.~maps
$q(y)$ back onto $p(x|\theta)$ (see again Theorem 2.12 in Ref.~\cite{villani_topics}).

The optimisation in the original problem in Equation \ref{eq:monge} is restricted to a very
specific set of functions $\Tth$, namely those for which $p'(x'|\theta)\equiv q(y)$.
On the other hand, the dual formulation in Equation \ref{eq:dual} is subject to the much more
generic (and manageable) requirement that $\fth(y;\theta)$ be a convex function.
This provides an avenue towards solving the dual problem numerically, and then reconstructing
the optimal transport map by virtue of Equation \ref{eq:gradient_reconstruction}.

\subsection{A more practical form}

To express Equation \ref{eq:dual} in the form of a functional optimisation problem, we make use
of a property of the convex conjugate.
The argument $\hat{y}$ that maximises the right-hand side of Equation \ref{eq:legendre_transform}
is given by the gradient of the transformed function, i.e.~$\hat{y}=\grad_x \fth^*(x;\theta)$.
We thus have $\fth^*(x;\theta) = x \cdot \grad_x \fth^*(x;\theta) - \fth(\grad_x\fth^*(x;\theta); \theta)$ and the convex conjugate
$\fth^*$ can also be computed via
\beq
\fth^*(x;\theta) = \argsup_{\gth \in \cvx(X)} x \cdot \grad_x \gth(x;\theta) - \fth(\grad_x \gth(x;\theta); \theta).
\eeq

As noted in Ref.~\cite{OT_with_convex_NNs}, the dual problem may then be rewritten as
\begin{align}
  \label{eq:minimax}
  (\hat\fth, \hat\gth) = \arg &\inf_{\fth \in \cvx(Y)} \sup_{\gth \in \cvx(X)} \int dy\,\, q(y) \, \fth(y;\theta) \nonumber \\
  &+ \int dx\,\, p(x|\theta) \, \left[ x \cdot \grad_x \gth(x;\theta) - \fth(\grad_x \gth(x;\theta); \theta) \right].
\end{align}
This minimax problem can now be solved efficiently by standard optimisation methods.
Since $\hat\gth(x;\theta) \equiv \hat\fth^*(x;\theta)$, the optimal transport function is then given by
\beq
\hat\Tth(x;\theta) = \grad_x \hat\gth(x;\theta).
\label{eq:transport_function_practical}
\eeq

\subsection{Optimal transport for one-dimensional distributions}
The results stated above remain true, of course, in the special case of one-dimensional probability
distributions.
However, the solution for this situation has very peculiar characteristics, and thus deserves a special mention.

In one dimension, gradients of convex functions---and thus the transport function $\hat\Tth(x;\theta)$---are \emph{nondecreasing},
i.e.~the calibration never corrects the (now scalar-valued) events in different directions.
Moreover, $\hat\Tth(x;\theta)$ can be expressed explicitly as
\beq
\hat\Tth(x;\theta) = Q^{-1} \circ P(x|\theta),
\label{eq:OT_1d}
\eeq
where $P(x|\theta)$ and $Q(y)$ are the cumulative distribution functions corresponding to the densities $p(x|\theta)$ and
$q(y)$. A rigorous derivation of Equation \ref{eq:OT_1d} is given in Section 2.2 of Ref.~\cite{villani_topics}.
This simply states that applying quantile matching (a very widely used technique) is equivalent to our (much more general)
philosophy of calibration through optimal transport.

In the case where both distributions are Gaussians, i.e.~$p(\cdot|\theta) = \No(\mu_1(\theta), \sigma_1(\theta))$ and
$q(\cdot) = \No(\mu_2, \sigma_2)$, the transport function results in the well-known expression
\beq
\hat{\Tth}(x;\theta) = \frac{x-\mu_1(\theta)}{\sigma_1(\theta)} \cdot \sigma_2 + \mu_2.
\eeq

\section{Practical implementation of optimal transport}
\label{sec:implementation}

To solve the minimax problem in Equation~\ref{eq:minimax}, we replace the integrals
by their respective batched estimates using the dataset $\mathcal{X}$,
produced by the uncalibrated simulator, and the calibration dataset $\mathcal{Y}$.
Choosing a suitable parametrisation for the functions $f(y;\theta)$ and
$g(x;\theta)$ then turns the minimax problem into a tractable, finite-dimensional
optimisation problem.

In this section, we focus on the case where all simulation parameters $\theta$ are
continuous. They are typically associated with a prior distribution $p(\theta)$.
The dataset $\mathcal{X}$ extracted from the simulator then consists
of (weighted) samples of the form $(\{x_i, w_i, \theta_i\})_{i=1\ldots N}$,
obtained from the uncalibrated distribution $p(x|\theta) p(\theta)$.
Clearly, the calibration dataset $\mathcal{Y}$ does not intrinsically depend on
$\theta$, and so consists of samples $\{(y_j, \theta_j)\}_{j=1\ldots M}$ drawn
from the product distribution $q(y) p(\theta)$.

We parametrise the functions $\fth(y;\theta)$ and $\gth(x;\theta)$ with two neural
networks with parameters $\phi$ and $\psi$, respectively.
The network architecture (explained below) is chosen such that the parametrisations
$\fth_{\phi}(y;\theta)$ and $\gth_{\psi}(x;\theta)$ express manifestly convex
functions w.r.t.~the arguments $y$ and $x$ (as required by Equation \ref{eq:minimax}),
but that the dependency w.r.t.~the simulation parameters $\theta$ can be arbitrary.
This choice reflects the fact that, for continuous simulation parameters,
the uncalibrated distribution $p(x|\theta)$---and also the functions
$f(y;\theta)$ and $g(x;\theta)$---are expected to depend smoothly on $\theta$.

This gives rise to the loss functional in terms of the parameters $\phi$ and $\psi$,
\beq
\mathcal{L}_{\mathrm{trans.}}[\phi, \psi] = \frac{1}{M_b} \sum_{j=1}^{M_b} \, \fth_\phi(y_j;\theta_j) + \frac{1}{\sum_{i} w_i} \sum_{i=1}^{N_b} w_i \left[ x_i \cdot \grad_x \gth_\psi(x_i;\theta_i) - \fth_\phi(\grad_x\gth_\psi(x_i; \theta_i); \theta_i) \right],
\label{eq:loss_function_discretised}
\eeq
where $N_b$ and $M_b$ are the respective batch sizes.
In terms of Equation \ref{eq:loss_function_discretised}, the optimisation problem becomes
\beq
(\hat\phi, \hat\psi) = \arg \inf_{\phi} \sup_{\psi} \mathcal{L}_{\mathrm{trans.}}[\phi, \psi],
\eeq
and the optimal transport function is approximated by
\beq
\hat{T}(x;\theta) \approx \grad_x g_{\hat{\psi}}(x;\theta).
\eeq

\subsection{Network architecture}
\label{subsec:network_architecture}

In Ref.~\cite{ICNN}, a network architecture is constructed (referred to as
``partially input convex neural networks'', or PICNNs) which meets the
requirements set out above for the parametrisations of $f(y;\theta)$ and $g(x;\theta)$.
The employed strategy consists in first finding an architecture that ensures
convexity w.r.t.~the function arguments, and then adding on top of it the most
general structure to cover the dependency w.r.t.~the parameters $\theta$.

\begin{figure}[tph]
  \centering
  \includegraphics[width=14cm]{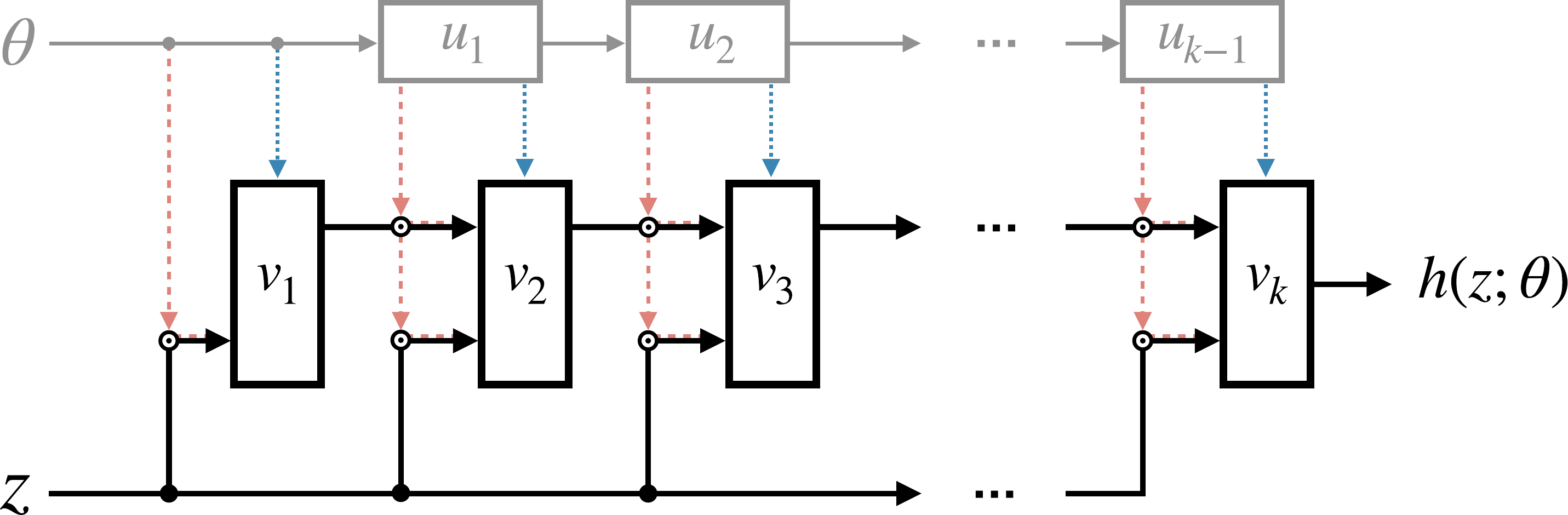}
  \caption{Architecture of a neural network implementing a function $\hth(z;\theta)$, which is convex w.r.t.~the argument $z$, but not necessarily w.r.t.~the parameters $\theta$. See the main text for more detailed explanations. This architecture is used here to parametrise the functions $\fth(x;\theta)$ and $\gth(y;\theta)$.}
  \label{fig:PICNN_architecture}
\end{figure}

The full architecture of a network implementing a generic function $\hth(z;\theta)$ (used
to represent both $\gth(x;\theta)$ and $\fth(y;\theta)$) is visualised in Figure
\ref{fig:PICNN_architecture}. It is defined through the recursion relations
\begin{align*}
  v_{i+1} = g_i^{(v)}\Biggl( &W_i^{(v)}\left[v_i \odot \textcolor{pastelorange}{\left(W_i^{(vu)}u_i + b_i^{(vu)} \right)_+} \right] \\
  + &W_i^{(z)}\left[z \odot \textcolor{pastelorange}{\left(W_i^{(zu)}u_i + b_i^{(zu)}\right)} \right] \\
  + &\textcolor{pastelblue}{W_i^{(u)}u_i} + b_i^{(v)} \Biggr),\quad v_0 = 0, \quad h(z;\theta) = v_k,
\end{align*}
and
\beqs
\quad u_{i+1} = g_i^{(u)}\left(W_i^{(u)}u_i + b_i^{(u)}\right), \quad u_0 = \theta.
\eeqs

The key observation made in Ref.~\cite{ICNN} is that a positive linear combination of
convex functions remains convex, as does the composition of a convex function with
a convex nondecreasing function.
This can be realised with a setup akin to a conventional feedforward neural
network with hidden nodes $v_i$, provided that the weights $W_i^{(v)}$ remain
nonnegative at all times. No restrictions exist on the biases $b_i^{(v)}$.
Additional ``bypass'' connections with weights $W_i^{(z)}$ route the inputs to
each hidden layer, which increases the expressiveness of the network while
maintaining its convexity properties.

The most general dependence on the parameters $\theta$ can be implemented by adding
all terms that are not in conflict with the convexity requirements. This allows an
auxiliary feedforward structure with $k-1$ layers.
Its hidden layers $u_i$ can be coupled to the main branch by decorating them onto
the $v_i$ (through weights $W_i^{(vu)}$ and biases $b_i^{(vu)}$) as well as the
bypass connections (through $W_i^{(zu)}$ and $b_i^{(zu)}$).
This is performed by means of the Hadamard (element-wise) product, denoted as $\odot$.
To ensure that this does not compromise the nonnegativity of the weights $W_i^{(v)}$,
the decorated term itself must be nonnegative, indicated by the notation $\left(\cdot\right)_+$.
The $u_i$ can also enter directly as an additive term with weights $W_i^{(u)}$.

As mentioned above, the activation functions $g_i^{(v)}(s)$ must be convex
nondecreasing. No restrictions are posed on the activations of the auxiliary path, $g_i^{(u)}$.
According to Equation \ref{eq:transport_function_practical}, the transport function
is determined by the \emph{gradient} of $\gth_\psi$. To ensure a smooth transport, the
$g_i^{(v)}$ should therefore also have a smooth first derivative.
Using a parabolic function to smoothly connect the two line segments in a leaky
ReLU gives rise to an activation function $\act(s)$ which satisfies all requirements,
\beq
  \act(s) =
    \begin{cases*}
      b s & $s < 0$ \\
      b s + \frac{c - b}{2} s^2 & $0 \leq s \leq 1$ \\
      \frac{b - c}{2} + c s & $1 < s$,
    \end{cases*}
\eeq
where $b$ and $c$ correspond to $\act'(s)$ for $s < 0$ and $1 < s$, respectively.
For this study we have chosen $b = 0.05$, $c = 1$, although these parameters
were not optimised.

\subsection{Training methodology}

During training, the parameters of the networks expressing $f_\phi$ and $g_\psi$
are updated iteratively, similar to the joint training of the generator and
adversary in Generative Adversarial Networks~\cite{gans}.
According to Equation \ref{eq:minimax}, the training proceeds schematically as
\beq
\label{eq:update}
\phi \leftarrow \phi - \eta_f \grad_\phi \mathcal{L}, \qquad \psi \leftarrow \psi + \eta_g \grad_\psi \mathcal{L}.
\eeq
In practice, it can be beneficial to apply multiple update steps to $\psi$ for
each $\phi$ update, and/or to use a larger learning rate, $\eta_g > \eta_f$.

To ensure convexity of $\fth(y;\theta)$ and $\gth(x; \theta)$, the nonnegativity
of the corresponding weights $W_i^{(v)}$ must be maintained; to achieve this,
a suitable regularisation term is added to the training loss, and the total
loss function becomes
\beq
\label{eq:fullloss}
\mathcal{L}[\phi, \psi] = \mathcal{L}_{\mathrm{trans.}}[\phi, \psi] +
\sum_{W_i^{(v)} \in \phi, \psi} \mathrm{max}(-W_i^{(v)}, 0).
\eeq
Ensuring the nonnegativity of these weights in a ``soft'' manner by means of
these regularisation terms was found to lead to a more stable training compared
to forcing all negative weights to zero after each update step. This is in line
with the observations made in Ref.~\cite{OT_with_convex_NNs}.

\section{Application in experimental particle physics}
\label{sec:results}

We explore three examples with increasing complexity in order to illustrate the
effectiveness of our method in a situation inspired by experimental particle
physics.
While these examples are very simple and therefore easily visualised and
understood, we emphasise that they are arbitrary and that the method described
previously can be applied to similar problems of greater complexity and higher
dimensionality.

\subsection{Calibrating the signal prediction}
\label{subsec:signal}
In experimental particle physics, it is common to study the distributions of
physically motivated event variables.
One such variable is the invariant mass of a system of particles, for which
the physical process of interest, the ``signal'', typically produces a distinct
peak in the observed data, corresponding to a short-lived particle created in the
collision.

Here we consider a signal process characterised by a density that is the sum of
two normal distributions: a ``core'' with $\mu = 0.1$ and $\sigma = 0.1$ and a
``tail'' with identical mean, but $\sigma = 0.25$.
In our example, we suppose the core and the tail distributions come in equal proportions.
This corresponds to the common scenario in which there are several detector effects
that determine the overall resolution on the observed mass variable $y$.

We assume that these resolution effects are not well-known \emph{a priori} and that
they are thus not modelled by the simulation: it predicts the uncalibrated mass
variable $x$ to follow a single normal distribution with $\mu = -0.5$ and $\sigma = 0.25$.
It is then the objective of the transport function $x' = \hat{T}(x)$ to rectify the
discrepancy in the shape of the peak between the simulator prediction and the observed
data.
For this simple example, we take the simulation to have no additional parameters,
i.e.~the argument $\theta$ is absent.

To construct the transport function, we follow the procedure described in
Section~\ref{sec:implementation}.
The architecture of the networks parametrising $f_\phi(y)$ and $g_\psi(x)$ is that of
Figure \ref{fig:PICNN_architecture}.
We use relatively small networks, with three hidden layers of 32 $v_i$ nodes
each, using the PyTorch Deep Learning Library~\cite{torch}.
The $u_i$ nodes, corresponding to non-convex inputs, are not required in this case.

\begin{figure}

  \subfigure[][$f_{\hat \phi}(y)$]{
    \includegraphics[width=8cm]{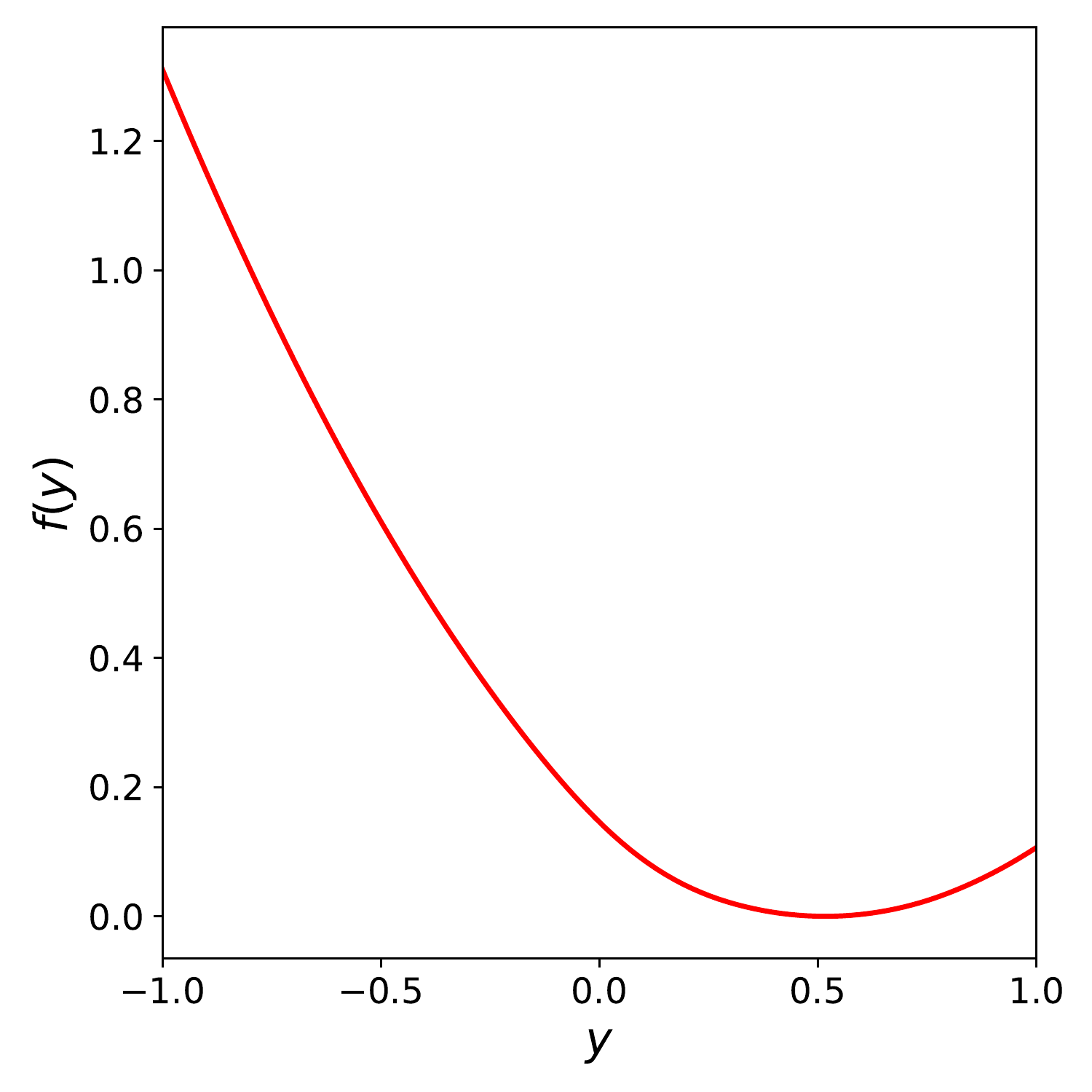}
  }
  \subfigure[][$g_{\hat \psi}(x)$]{
    \includegraphics[width=8cm]{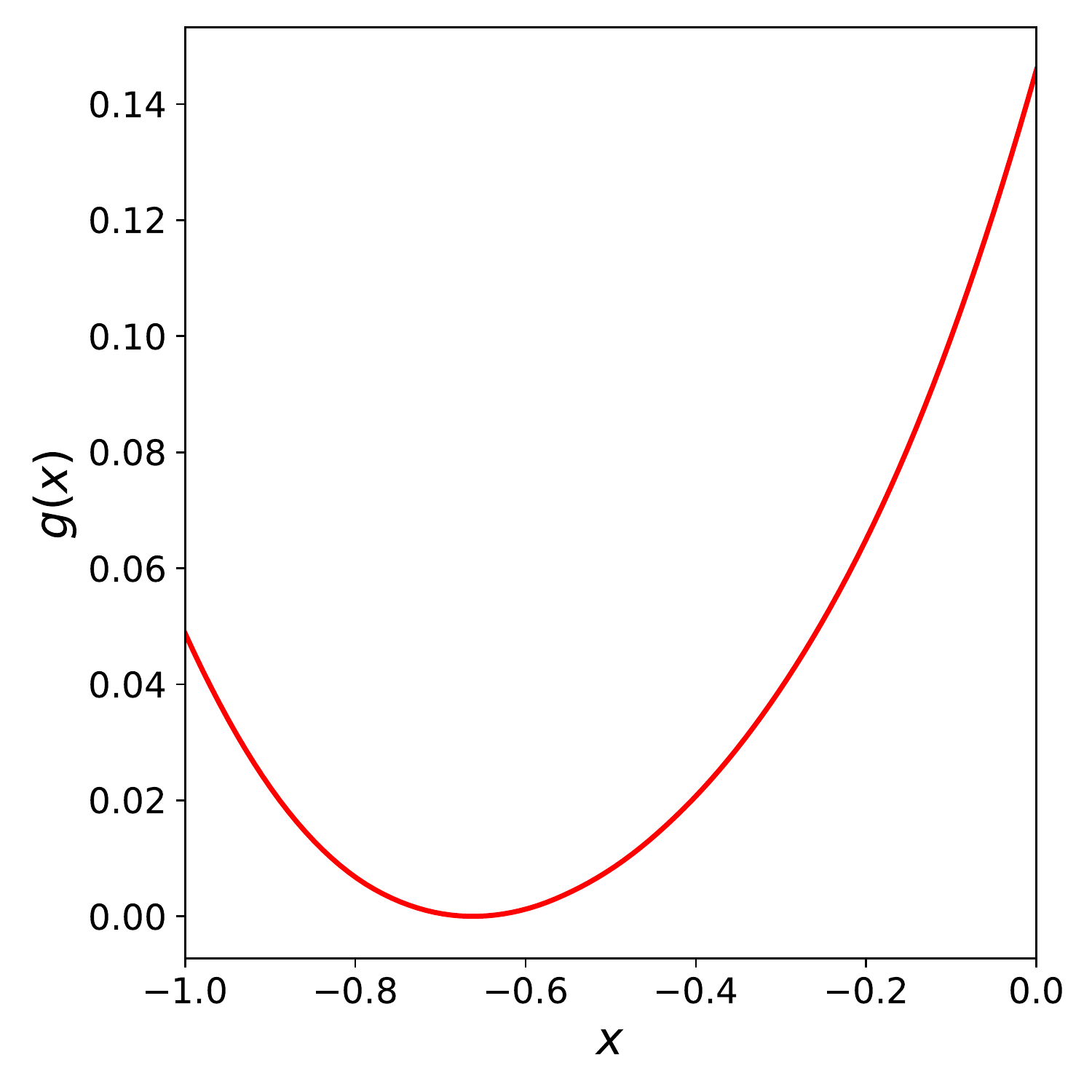}
  } \\
  \subfigure[][$T(x) = \grad_x\, g_{\hat \psi}(x)$]{
    \includegraphics[width=8cm]{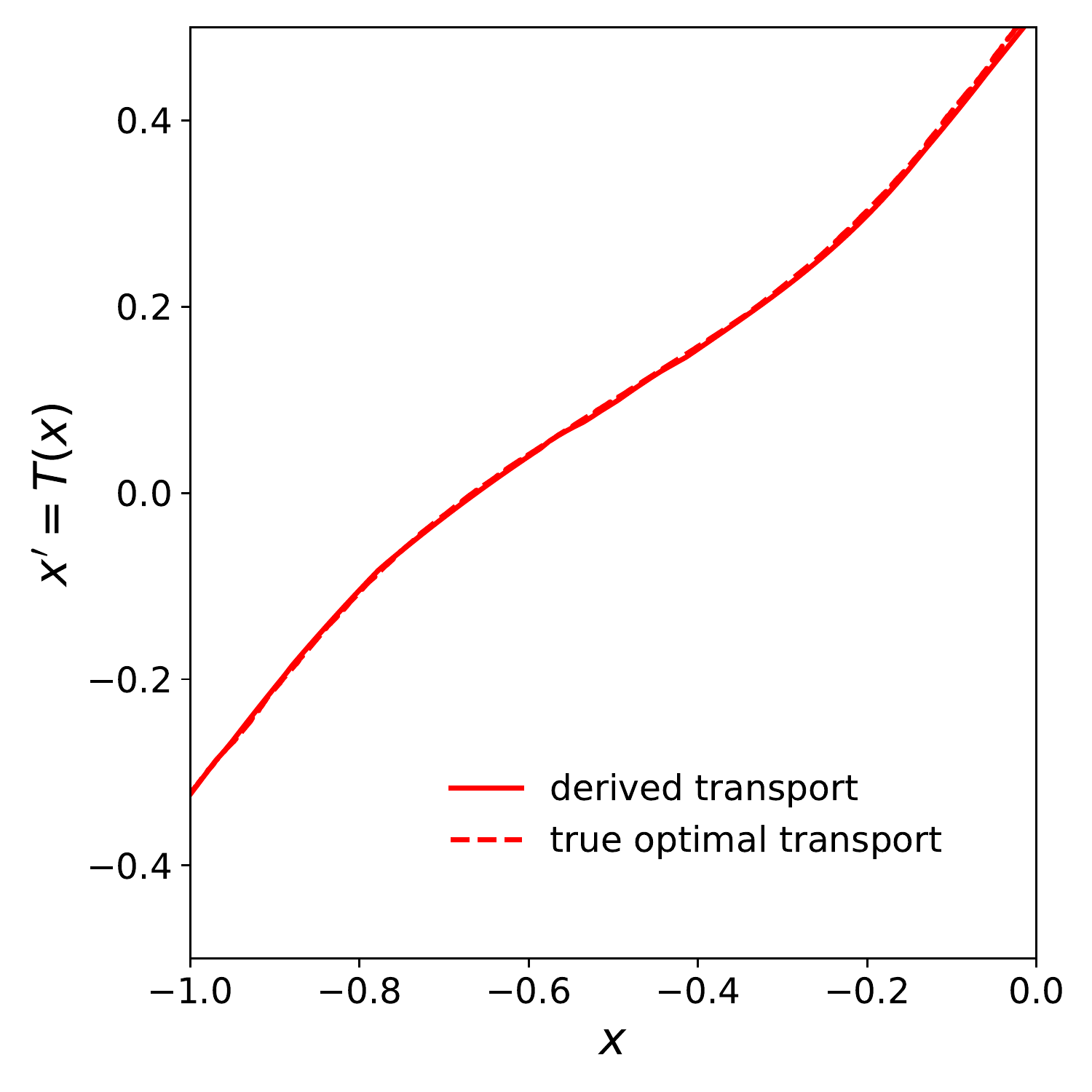}
  }
  \subfigure[][calibration data, uncalibrated and calibrated simulation]{
    \includegraphics[width=8cm]{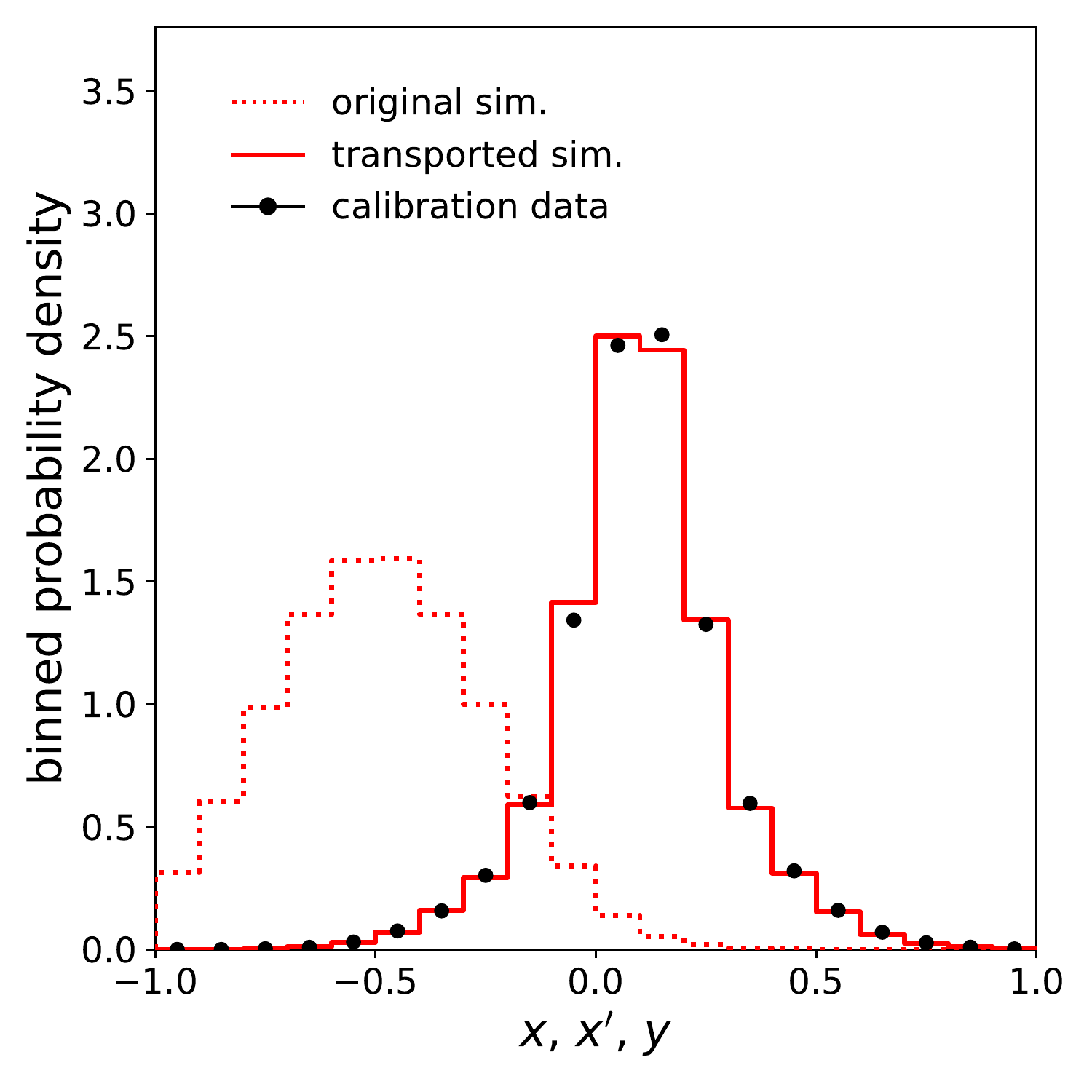}
  }
  \caption{
  The functions (a) $f_{\hat \phi}(y)$, (b) $g_{\hat \psi}(x)$, and (c) the resulting
  transport $\hat T(x) = \grad_x\, g_{\hat \psi}(x)$ after the training
  procedure for a signal-only scenario.
  The original simulated prediction, $p(x)$, and transported prediction,
  $p'(x')$, are compared to the calibration data distribution, $q(y)$, in (d), with
  closure observed at the 5\% level in each bin after the calibration is applied.
  }
  \label{fig:sigtrans}
\end{figure}

Figure~\ref{fig:sigtrans} shows the transport function resulting from following
this procedure, as well as a comparison of the data and simulated distributions
before and after the calibration is applied. 
Excellent closure is obtained after about 30 minutes of training on a laptop
computer using batches of $N_b = M_b = 256$ samples, the Adam
optimiser~\cite{adam}, and learning rates of $\eta_f = \eta_g = 10^{-4}$; in
this case we update the parameters $\psi$ ten times for each $\phi$ update.
These hyperparameters were chosen by performing a coarse scan of batch size
(from 16 to 1024), learning rate (from $10^{-3}$ to $10^{-5}$), and number of
nodes per layer (from 8 to 64), and choosing the parameters with the smallest
binned $KL$-divergence between transported simulation and calibration data.

\subsection{Calibration in the presence of background processes}
\label{subsec:background}

It is not typically the case that a sample of events is available which is
perfectly pure in the signal process.
Instead, contributions from \emph{backgrounds} are also important, i.e.~physical
processes that are not immediately of interest, but also contribute to the
collection of studied events, and that must thus be understood.
In many cases, background processes lead to distributions over the observed mass variable
$y$ that are rather broad and smooth, and do not contain any sharp features.

To illustrate how backgrounds can be accommodated in our formalism, we now
suppose that the observed data are actually drawn from two distributions: the
signal events still come from the double-Gaussian described in the example in Section
\ref{subsec:signal}, but 50\% of all events originate from a background process
for which $y$ follows a normal distribution with $\mu = -0.5$ and $\sigma = 1$.

We suppose that the simulation of the backgrounds is already accurate, and does
not require calibration.
The objective then becomes to construct a transport function $\hat T(x)$ that
ensures that the simulation of the signal is properly calibrated.

This means that we require $\hat T(x) = x$, whenever $x$ corresponds to a
simulated background event. To achieve this, we define
\beq
  g_\psi (x) =
    \begin{cases*}
      g^{\mathrm{sig}}_\psi(x) & \text{for signal events} \\
      \frac{1}{2} x^2 & \text{for background events},
    \end{cases*}
    \label{eq:sig_only_transport}
\eeq
noting that $\grad_x \frac{1}{2}x^2= x$ for background events, as required.
The function $g^{\mathrm{sig}}_\psi(x)$ remains unspecified and is adjusted
during the training procedure in the usual way to yield a minimal modification
of the signal simulation that results in its calibration.

Equation \ref{eq:sig_only_transport} goes a rather direct route towards constructing
a transport function with the desired properties. Appendix \ref{sec:background_subtraction}
shows that this form of $g_\psi(x)$ is also formally correct, i.e.~it effectively
leads to a subtraction of the background prediction from the total distribution contained
in the calibration dataset in order to isolate and calibrate the signal.

\begin{figure}

  \subfigure[][$\hat{T}(x) = \grad_x\, g_{\hat \psi}(x)$]{
    \includegraphics[width=8cm]{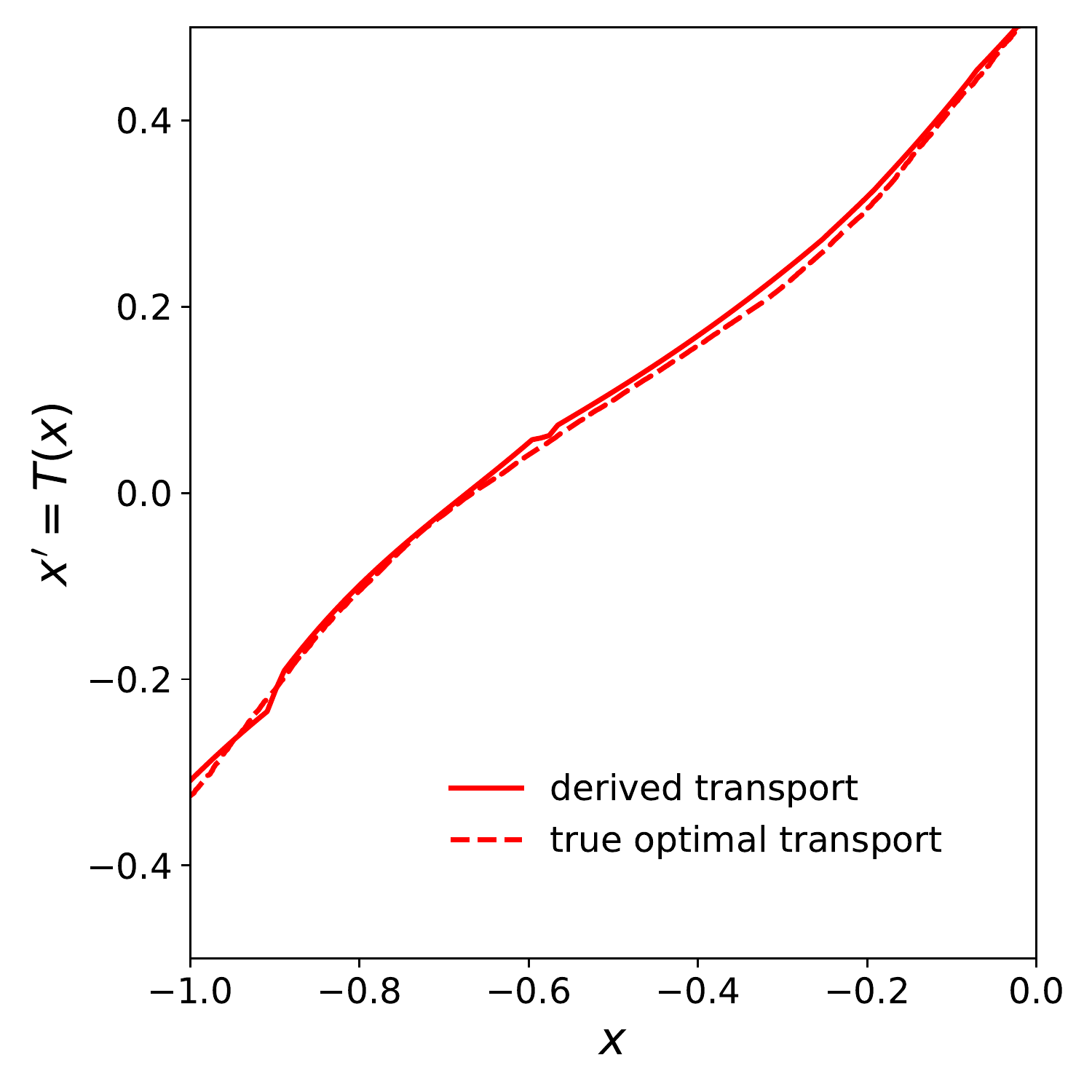}
  }
  \subfigure[][target, prediction, and transported distributions]{
    \includegraphics[width=8cm]{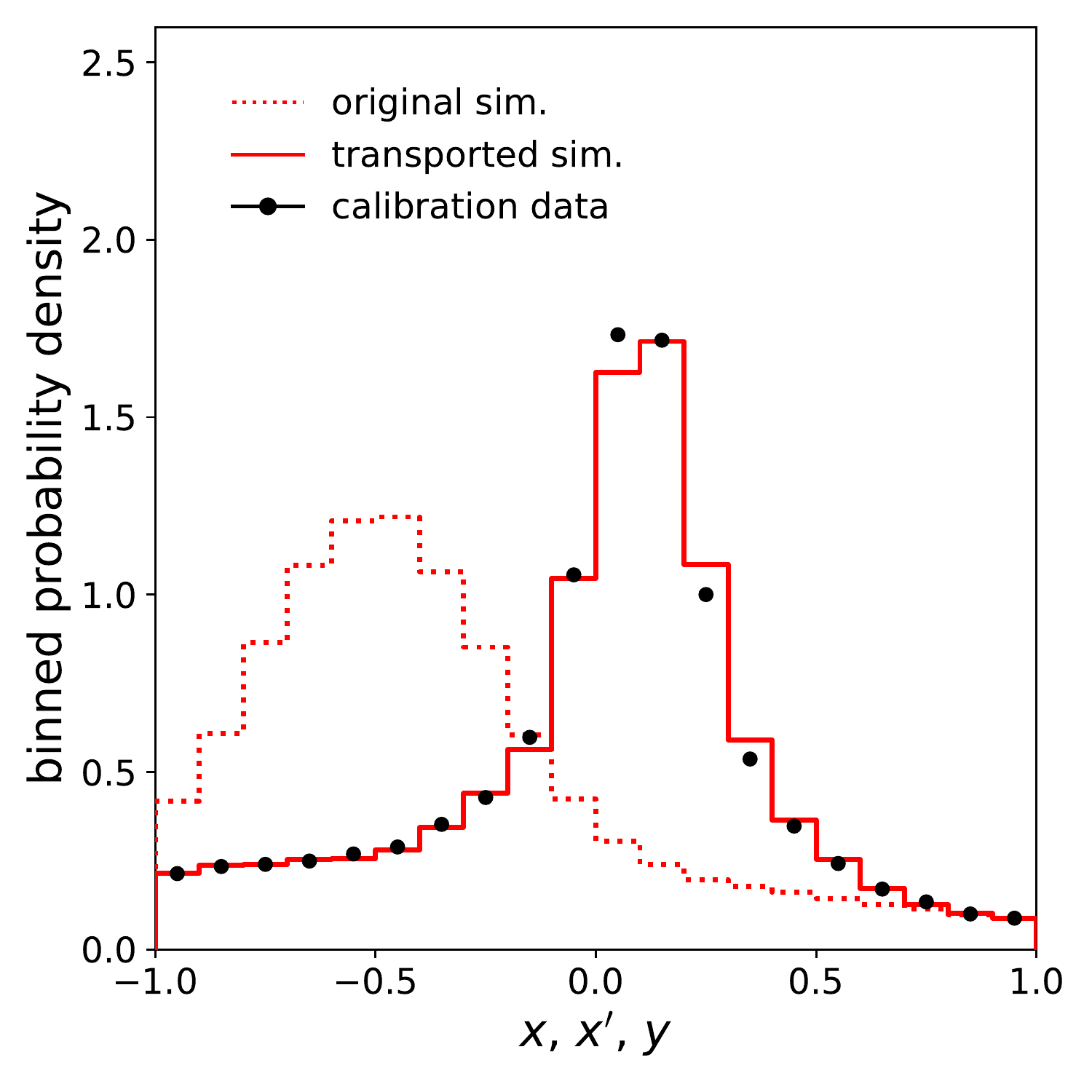}
  }
  \caption{
  (a) the transport function $\hat{T}(x) = \grad_x\, g_{\hat \psi}(x)$
  after the training procedure for a signal-plus-background scenario. 
  The original simulated prediction, $p(x)$, and transported prediction,
  $p'(x')$, are compared to the calibration data distribution, $q(y)$, in (d), with
  closure observed at the 5\% level in each bin after the calibration is applied.
  }
  \label{fig:sigbkgtrans}
\end{figure}

We proceed with the minimisation over the network parameters $\phi$ and $\psi$ in
the usual way to obtain $f_{\hat\phi}$ and $g_{\hat\psi}$.
Also the network architecture introduced in Section \ref{subsec:signal} remains
unchanged.
Figure~\ref{fig:sigbkgtrans} shows the signal transport function
$\grad_x \, g^{\mathrm{sig}}_{\hat \psi} (x)$ and comparisons between data and simulation
before and after the calibration is applied.
Excellent closure is again observed between observed data and the calibrated
simulator prediction.

\subsection{Systematic uncertainties}

We now introduce additional simulation parameters $\theta$, which control the exact
shape of the simulated signal distribution.
By virtue of their prior distribution $p(\theta)$, the simulation parameters give
rise to a notion of systematic uncertainties on the prediction of the uncalibrated
simulation.
In this context, these parameters are often referred to as ``nuisance parameters''
since they are not of particular (physics) interest, but they nevertheless affect the
simulation, and hence its calibration.

To illustrate the handling of systematic uncertainties in our calibration strategy,
we return again to our example from Section~\ref{subsec:background}.
We extend it by introducing a single new parameter, $\theta_0$, which alters the simulator
such that the predicted centroid and width of the signal distribution depend linearly
on it:
\begin{align*}
  \mu(\theta_0) &= -0.5 + 0.1\, \theta_0, \\
  \sigma(\theta_0) &= 0.25 + 0.1\, \theta_0.
\end{align*}

The functions $f_\phi(y;\theta)$ and $g_\psi(x;\theta)$ now make use of the full
neural network architecture shown in Section \ref{subsec:network_architecture}.
We assume the prior distribution $p(\theta_0)$ to correspond to a standard normal
distribution, as is common in experimental analyses~\cite{lhcstats}.
This imposes a Gaussian constraint on $\theta_0$, with $\theta_0 = 0$
corresponding to the ``nominal'' best-guess value of the parameter.
The training proceeds according to the loss function in Equation
\ref{eq:loss_function_discretised}, and during training we sample values of $\theta_0$ according
to a uniform distribution in the range  $(-2, 2)$.
With the Gaussian constraint, this ensures good behaviour at the 2$\sigma$
level; if stability is needed beyond $2\sigma$, this is achievable by extending
the training range of $\theta_0$. 

Due to the added complexity needed to obtain closure for a range of $\theta_0$,
we use slightly more complex networks to implement $f_\phi$ and $g_\psi$ in this case.
Using the nomenclature from Figure~\ref{fig:PICNN_architecture}, the networks
are composed of three hidden layers, with 64 $v_i$ nodes and 16 $u_i$ nodes in
each layer.

\begin{figure}

  \subfigure[][$\hat{T}(x;\theta_0) = \grad_x\, g_{\hat \psi}(x; \theta_0)$]{
    \includegraphics[width=8cm]{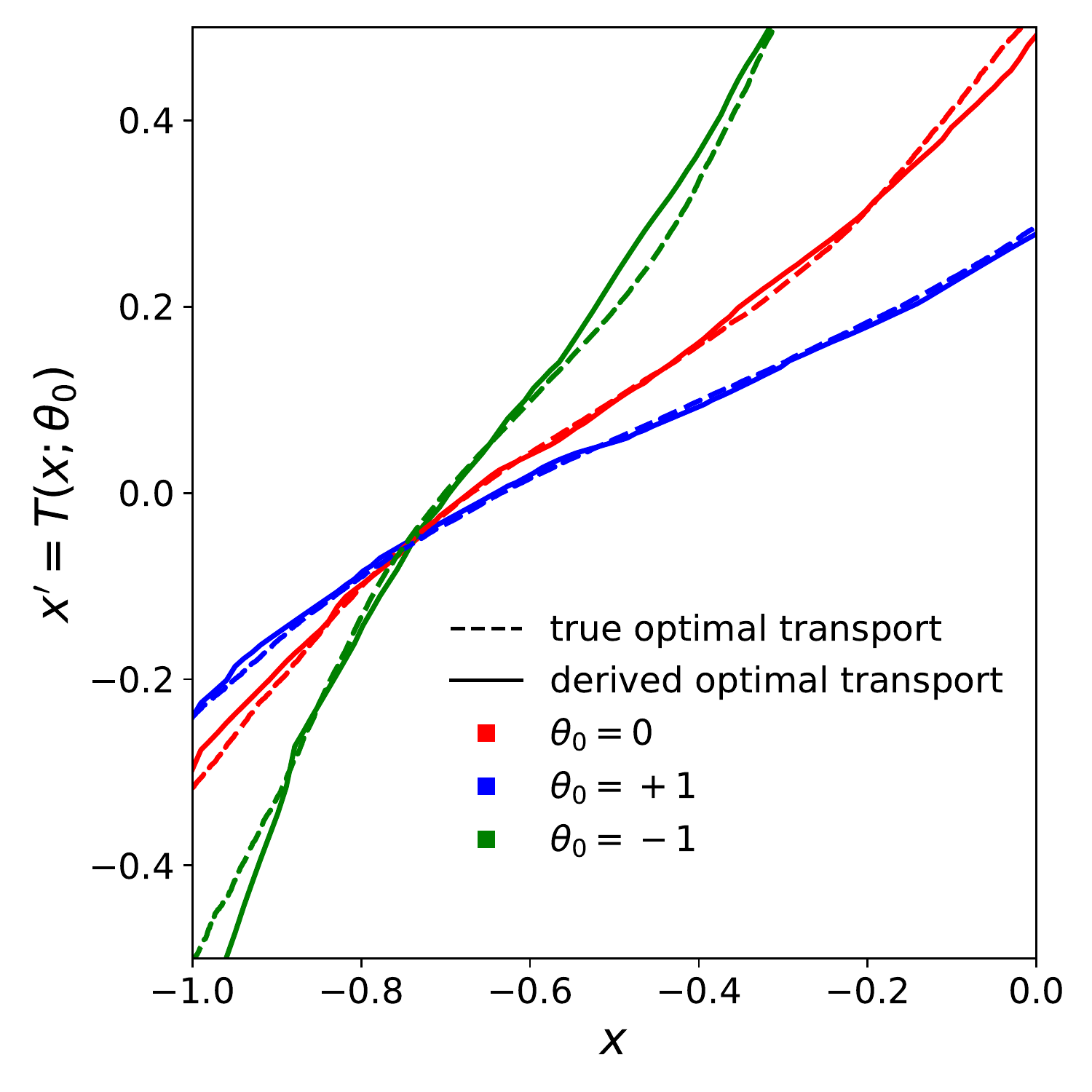}
  }
  \subfigure[][calibration data, uncalibrated and calibrated simulation]{
    \includegraphics[width=8cm]{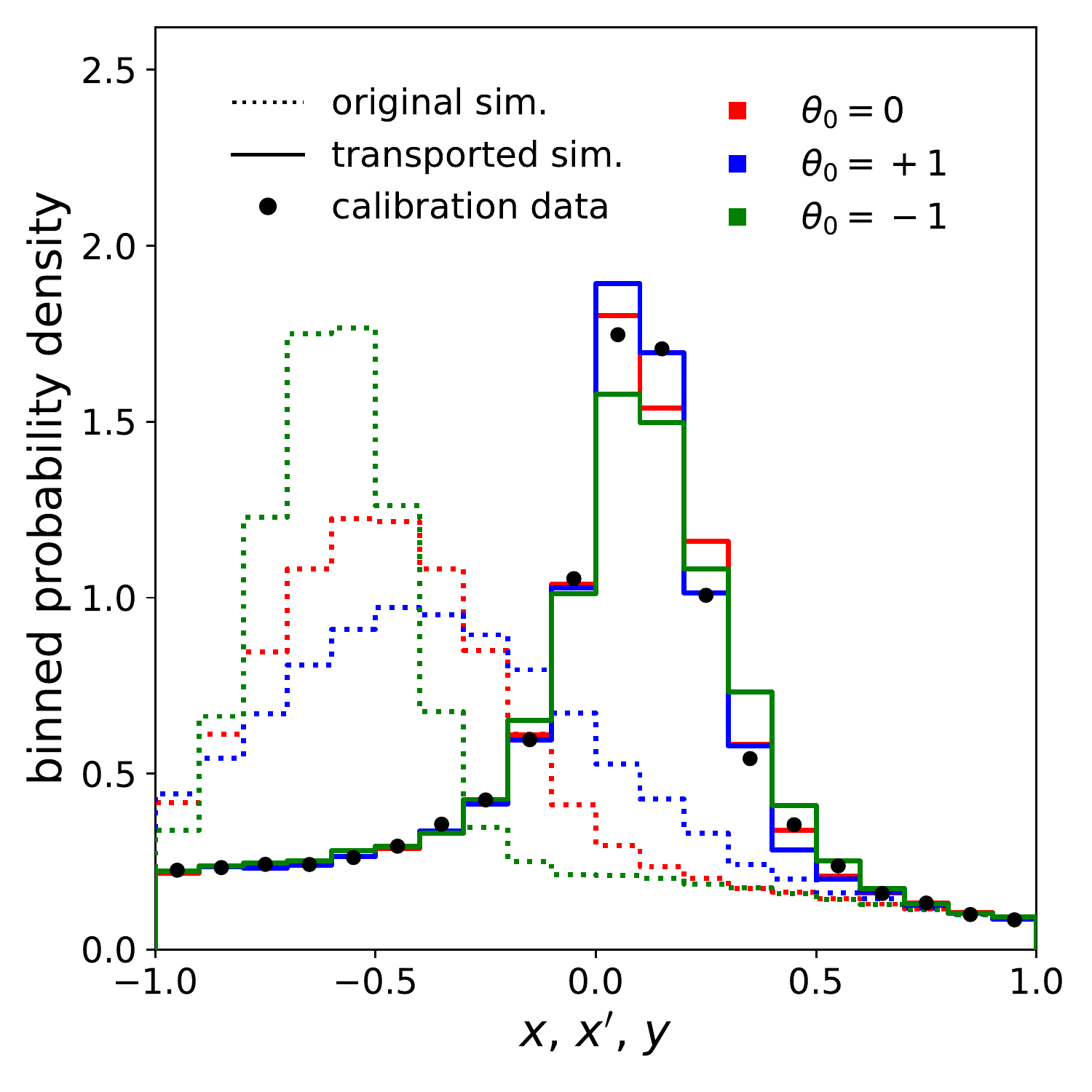}
  }
  \caption{
  (a) the transport function $T(x;\theta_0) = \grad_x\, g_{\hat \psi}(x;\theta_0)$
  after the training procedure for a signal-plus-background scenario. 
  The original simulated prediction, $p(x)$, and transported prediction,
  $p'(x')$, are compared to the target data distribution, $q(y)$, in (b), with
  closure at the 10\% level observed in each bin after the calibration is
  applied for each of the three $\theta_0$ points of interest.
  }
  \label{fig:sigbkgtranstheta0}
\end{figure}

The simulator and data distributions as well as the corresponding transport
functions and their dependence on $\theta_0$ are shown in
Figure~\ref{fig:sigbkgtranstheta0}.
Good agreement is achieved between the target dataset and transported simulation
for a wide range of $\theta_0$ values, and the resulting $\hat{T}(x;\theta_0)$ defines
a family of transport calibrations that can be used to evaluate the uncertainty on
the calibration $\hat{T}$ when applied in a signal region of interest.

\section{Conclusions and outlook}
\label{sec:conclusions}

The use of calibration data to correct details of stochastic simulators is an
integral component of the experimental physicist's toolkit.
We argue that the language of transportation theory is well-suited
for describing the adjustment of predictive models based on experimental
control samples. 
Calibrations derived using optimal transport methods are guaranteed to
minimally alter simulated predictions, which contain years of collected
domain knowledge, while faithfully reproducing relevant distributions observed
in calibration data.

We have presented a concrete implementation plan for obtaining calibration
transport maps, which are realised via feed-forward neural networks that are
straightforward to train on modern computing hardware. 
Illustrative examples were shown, demonstrating the capability of the
method and its ability to handle background processes and systematic uncertainties
on the simulated prediction.
Calibrations derived as transport maps and implemented as neural networks
overcome a number of shortcomings inherent to current methods, which are
less general, do not scale well into high-dimensions, or do not preserve some
important invariants such as the total number of predicted events.
Based on the generality of this approach and the fact that it depends on an
already well-established technology (neural networks), our method promises wide
applicability across experimental physics.

Optimal transport also provides a rigorous framework in which to generalise the
calibration method presented here.
In general, one can allow the simulator to produce data that populate a Riemannian
manifold \cite{OT_and_curvature, riemannian_convex}, on which a natural cost function
is given by the metric distance integrated along geodesics.
Taking the global structure of the data into account is important for the proper treatment
of components that are compact or periodic, e.g.~angular directions.
Furthermore, it can be useful to decouple the distribution that is being calibrated
(e.g.~the di-jet invariant mass) from the variable to which the calibration is applied
(e.g.~the transverse momenta of the jets).
In effect, this defines a relaxation of the original Monge problem in Equation \ref{eq:monge}.
It applies very naturally to situations where corrections should be applied directly to the
primary output of the simulator, but only the proper calibration of derived quantities
(possibly of lower dimension) is important.

\section*{Acknowledgements}

We thank Jose Clavijo Columbie, Marco Cristoforetti, and Judith Katzy for their
useful discussions, and Daniela Bortoletto for helpful feedback on this manuscript.
CP is supported by the Science and Technology Facilities Council (STFC) Particle
Physics Consolidated Grant ST/S000933/1.
PW acknowledges support by the STFC under grant ST/S505638/1 and through a
Buckee scholarship awarded by Merton College, Oxford.

\begin{appendices}

  \section{Calibrations in the presence of backgrounds: two equivalent ways}
\label{sec:background_subtraction}

We consider the calibration of the \emph{signal} in the presence of backgrounds.
The latter are assumed to be modelled correctly by the simulation, i.e.~do not
require a calibration.
The distribution delivered by the uncalibrated simulator is thus
\beq
p(x;\theta) = (1-\mu) \, p_{\mathrm{bkg}}(x) + \mu \, p_{\mathrm{sig}}(x;\theta),
\eeq
while the target distribution for the calibration is
\beq
q(y) = (1-\mu) \, p_{\mathrm{bkg}}(y) + \mu \, q_{\mathrm{sig}}(y).
\eeq
The coefficient $\mu$ corresponds to the signal fraction.

\subsection{One solution}
\label{sec:one_solution}

We thus want to construct a transport function that corrects the simulation of the signal, so that $p'_{\mathrm{sig}}(x';\theta)\equiv q_{\mathrm{sig}}(y)$.
According to Equation \ref{eq:minimax}, the corresponding transport function is the
solution of the minimax problem
\beq
(\hat{f}, \hat{g})=\arg\inf_{\fth} \sup_{\gth} \int dy\,\, q_{\mathrm{sig}}(y) \, \fth(y;\theta) + \int dx\,\, p_{\mathrm{sig}}(x|\theta) \, \left[ x \cdot \grad_x \gth(x;\theta) - \fth(\grad_x \gth(x;\theta);\theta) \right],
\label{eq:signal_transport}
\eeq
where, as usual, $f\in\cvx(Y)$ and $g\in\cvx(X)$. To make the following manipulations
more concise, we use the shortcut $\expt_{h}[k(x)]$ to denote the expectation of
$k(x)$ under the density $h(x)$, i.e.~$\expt_{h}[k(x)] = \int dx\, h(x) k(x)$.

To express Equation \ref{eq:signal_transport} in terms of the densities $q(y)$,
$p_{\mathrm{bkg}}(y)$ and $p_{\mathrm{sig}}(x;\theta)$, we make use of the relation
\beqs
\expt_{q_{\mathrm{sig}}} = \frac{1}{\mu} \expt_{q} - \frac{1-\mu}{\mu} \expt_{p_{\mathrm{bkg}}},
\eeqs
which follows from the identity
\beqs
q_{\mathrm{sig}}(y) = \frac{1}{\mu} q(y) - \frac{1-\mu}{\mu} p_{\mathrm{bkg}}(y).
\eeqs
With this, we find
\begin{align*}
  (\hat{f}, \hat{g}) &= \arg\inf_{\fth} \sup_{\gth} \frac{1}{\mu} \expt_q\left[f(y;\theta)\right] - \frac{1-\mu}{\mu} \expt_{p_{\mathrm{bkg}}}\left[f(y;\theta) \right] + \expt_{p_{\mathrm{sig}}}\left[ x \cdot \nabla_x g(x;\theta) - f(\nabla_x g(x;\theta); \theta)\right]\\
  & \equiv \arg\inf_{\fth} \sup_{\gth} \expt_q\left[f(y;\theta)\right] - (1-\mu) \expt_{p_{\mathrm{bkg}}}\left[f(y;\theta) \right] + \mu \expt_{p_{\mathrm{sig}}}\left[ x \cdot \nabla_x g(x;\theta) - f(\nabla_x g(x;\theta);\theta)\right].
\end{align*}

\subsection{The same solution again}

Alternatively, we can start from the original transport between $p(x;\theta)$ and $q(y)$,
\beq
(\hat{f}, \hat{g}) = \arg\inf_{\fth} \sup_{\gth} \expt_q\left[f(y;\theta)\right] + \expt_p\left[ x \cdot \nabla_x g(x;\theta) - f(\nabla_x g(x;\theta);\theta)\right]
\eeq
and insert the parametrisation from Equation \ref{eq:sig_only_transport} for $g(x;\theta)$, i.e.~demand that $g(x;\theta) = \frac{x^2}{2}$ whenever it is evaluated on a background event.

The above minimax problem then turns into
\begin{align*}
  (\hat{f}, \hat{g}) = \arg\inf_{\fth} \sup_{\gth} \expt_q\left[f(y;\theta)\right] + (1-\mu) \expt_{p_{\mathrm{bkg}}}&\left[y \cdot y - f(y; \theta)\right]\\
  + \mu \expt_{p_{\mathrm{sig}}}&\left[x \cdot \nabla_x g(x;\theta) - f(\nabla_x g(x;\theta);\theta)\right].
\end{align*}
Dropping the term $(1-\mu) \expt_{p_{\mathrm{bkg}}}\left[y \cdot y\right]$, which is
independent of $f$ and $g$, we get,
\beqs
(\hat{f}, \hat{g}) = \arg\inf_{\fth} \sup_{\gth} \expt_q\left[f(y;\theta)\right] - (1-\mu) \expt_{p_{\mathrm{bkg}}}\left[f(y;\theta)\right] + \mu \expt_{p_{\mathrm{sig}}}\left[ x \cdot \nabla_x g(x;\theta) - f(\nabla_x g(x;\theta);\theta)\right],
\eeqs
which is identical to the result from Section \ref{sec:one_solution}.

\end{appendices}

\end{document}